# Direct microscopic evidence of shear induced graphitization of ultrananocrystalline diamond films

K. Ganesan [1,2,*], Revati Rani [1], Tom Mathews [1,2], S. Dhara [1,2]

[1] *Surface and Nanoscience Division, Indira Gandhi Centre for Atomic Research, Kalpakkam-603102, India*

[2] *Homi Bhabha National Institute, Indira Gandhi Centre for Atomic Research, Kalpakkam-603102, India*

## Abstract

The origin of ultralow friction and high wear resistance in ultrananocrystalline diamond (UNCD) films is still under active debate because of the perplexed tribochemistry at the sliding interface. Herein, we report a comparative study on surface topography and nanoscale friction of tribofilms, in wear tracks of two sets of UNCD films having different structural characteristics. Despite both the films display ultralow coefficient of friction, the UNCD films grown under Ar atmosphere ($UNCD_{Ar}$) exhibit a high wear resistance while the wear rate is higher for the films grown in $N_2$ ($UNCD_N$). Frictional force microscopic (FFM) investigations clearly reveal the manifestation of shear induced graphitization on both the films. However, the wear track of $UNCD_{Ar}$ films have a large network of a few layer graphene (FLG) structures over the amorphous carbon tribofilms while only isolated clusters of FLG structures are present in the wear track of $UNCD_N$ films. Here, we demonstrate the direct micro-/nanoscopic evidence for the formation of large network of ~ 0.8 - 6 nm thick FLG structures, as a consequence of shear induced graphitization and discuss their decisive role in ultralow friction and wear.

[*] Corresponding author. Tel : +91-44-27480500 Extn: 21669. Email : kganesan@igcar.gov.in ( K. Ganesan)

# 1. Introduction

Friction and wear are one of the major challenges in mechanical assembly systems and they lead to significant energy loss and frequent device failure. The materials with anti-wear and ultralow friction characteristics are the most desirable in mechanical engineering applications where the carbon based materials have potentials to play a major role. Since ultralow friction and negligible wear are demonstrated in ultrananocrystalline diamond (UNCD) and diamond-like carbon (DLC) films at ambient conditions, the research interest is ever increasing drastically over the recent past years [1,2,11–20,3,21–30,4,31–34,5–10]. The formation of tribofilms which occurs due to the shear induced graphitization is one of the crucial factors for achieving ultralow friction in carbon materials [5,10,20–26,35]. Especially when the tribofilm consists of layered structures e.g. a few layer graphene (FLG), the friction becomes vanishingly low [7,31–33,36]. Since the 2D layered materials have outstanding structural, chemical, physical and mechanical properties due to the strong in-plane interactions and very weak van der Waals out of plane interactions, these materials become highly attractive for tribological applications, apart from the well demonstrated novel electronic and optoelectronic applications [37,38]. Apart from graphene layers, carbon/graphene quantum dots, graphene oxide, reduced graphene oxide, nanodiamonds and other 2D materials such as hexagonal BN, $MoS_2$, $NbSe_2$ and $WS_2$ have shown remarkably low friction and wear that encourages to use them as solid lubricants in tribology [13,38–41].

There are numerous reports available in the literature on the factors affecting the ultralow friction and wear of UNCD and DLC films. Among them, the surface roughness of tribo-contacts, contact pressure, tribo-environment, crystalline orientation, structural defects, mechanical properties, and surface chemistry – passivation by functionalization or the formation of secondary phase tribofilms due to tribo-chemical reactions at the sliding interface play significant role on tribological properties [14–20]. The chemical and microstructure analysis of the tribofilm can provide invaluable information about the tribochemical reactions at the interface and the wear mechanism. Despite significant knowledge gained over the period, the microscopic origin of ultralow friction and wear in UNCD films are still not well understood. The most acceptable postulates on ultralow friction and wear in UNCD and DLC are either due to surface passivation or rehybridization. Of them, the surface passivation is reasonably explained on the superlubricity in UNCD and DLC films based on tribotests at different environments [3–5,21,22]. On the other hand the surface rehybridization, which occurs through the shear pressure leading to either amorphization or graphitization, is a complex phenomenon at atomic-scale with bond breaking and re-bonding at the sliding interface.

The shear induced graphitization is mostly verified experimentally by observation of $sp^2$ and $sp^1$ rich *a*-C in the tribofilms which are studied by several analytical tools such as Raman spectroscopy, high resolution transmission electron microscopy (HRTEM), electron energy loss spectroscopy (EELS), X-ray photoelectron spectroscopy (XPS) and near-edge X-ray absorption fine structures [10,18,20,22–26,35]. Recently, *in-situ* Raman tribometry has given tremendous support in understanding the tribochemical reactions at the sliding interface [10]. However, the obtained information is not limited to surface alone but includes a significant contribution from bulk and also, the spatial resolution is poor. While XPS is a well-known surface sensitive tool, it offers average information over a wide area and limits the knowledge at micro-/nano-scopic scale [4,18,20,23]. Moreover, surface-enhanced Raman spectroscopy (SERS) and tip-enhanced Raman spectroscopy (TERS) are well established as surface sensitive tools with much improved lateral resolution as compared to conventional Raman spectroscopy. Despite the SERS and TERS are well-established for detection of single molecular level and have high potential for intensive tribological analysis at molecular level, the usage of these techniques are very limited until now [42].

*In-situ* and post wear HRTEM and EELS studies had shown evidences on the rehybridization of UNCD into $sp^2$- and $sp^1$-rich amorphous carbon (*a*-C) embedded with small $sp^2$ crystallites of ~ 1 nm, fullerene-like structures and graphene nanoscroll which are attributed to the superlubricity [2,6,10,23,25–27,35]. However, there are no direct microscopic evidence of large size two dimensional (2D) graphene structures in the tribofilms. Also, molecular dynamic simulations by different groups had predicted superlubricity on UNCD and DLC films. Nevertheless, the superlubricity is achievable only when the formation of 2D graphene layers are considered at the sliding interface with minimum of two or more layers [8,24,28–30]. Based on the experimental assumptions and theoretical predictions, there are several reports on usage of a few layer graphene (FLG) structures as a solid lubricants and it has been demonstrated reduction in wear and friction on various types of tribo-couples including superlubricity on carbon materials [7,31–33,36]. Thus, 2D graphene layers are well proved for their effectiveness in protecting the surface and minimizing friction.

Even though the concept of 2D graphene layer formation on tribofilms through shear induced graphitization is well recognized by both experimental and theoretical simulations, it is surprising to note here that direct micro-/nanoscopic evidences for the formation of layered graphene structures on tribotrack are still lacking in literature [27,34]. This is mainly because of the challenges in distinguishing the atomically thin layers from rough wear track by the conventional microscopic techniques. Interestingly, atomic force microscope (AFM) can play a major role in studying the tribofilms since it can work from hundreds of micrometers down to atomic-scale. In addition, it also has several functional capabilities such as studying friction, elastic modulus, electrical and magnetic properties. Further, the frictional force microscope (FFM) has been explored in probing friction of carbon based materials and several other layered structures at nanometer-scale [36,39,43–45]. Since the coefficient of friction (CoF) is different for various carbonaceous materials, a combined mapping of surface topography and friction using FFM would provide information about the tribofilms in the wear track. Table 1 provides the performance comparison of different analytical techniques for surface analysis of materials. As evident from the Table 1, AFM based techniques have an edge over other techniques for surface analysis, especially for carbon materials.

Table 1. A table of performance comparison of surface analysis by different experimental techniques

| Technique / Parameter | Raman | SERS / TERS | XPS | HRTEM / EELS | AFM based techniques |
|---|---|---|---|---|---|
| **Lateral resolution** | ~ 1 μm | Down to ~ 10 nm | ~ 1 μm | ~ 0.05 nm | ~ 0.1 nm |
| **Max. probing lateral length** | ~ 1 μm to > 100 μm | ~ 1 μm to > 100 μm | ~ 1 μm to > 100 μm | ~ < 100 nm | 1 nm to ~ 100 μm |
| **Probing depth** | ~ 100 nm to >10 μm; Depends on the nature of specimen | ~ a few nanometers | ~ 4 nm | Thickness of the specimen, < 100 nm | Surface / subsurface / bulk : Depends on the mode of operation |
| **3D imaging** | Possible under confocal configuration; Poor z-scale resolution | Not applicable | Possible with depth profiling; Poor z-scale resolution | Not applicable | Yes; With high z-scale resolution ~ 0.1 nm |
| **Principle** | Light scattering | Light scattering | Photoelectric effect | Electron diffraction | Van der Waals force interaction |
| **Working environment** | Ambient / liquid / vacuum | Ambient / vacuum | vacuum | vacuum | Ambient / liquid / vacuum |
| **Sensitivity** | Depends on Raman scattering cross section | Depends on the nature of plasmonic nanostructures | Increases with atomic number | Increases with atomic number | Equal for all materials |
| **Properties measured** | Vibrational modes which provide the information about structure and chemical bonding | Vibrational modes which provide the information about structure and chemical bonding | Electronic structure; composition and chemical analysis | Crystal structure, electronic structure, composition and chemical analysis | Surface topography, electronic, opto-electronic, magnetic, friction and mechanical properties |
| **Ease of operation** | Easy | Easy | Difficult | Difficult | Easy |
| **Operation cost** | Moderate | Moderate | Expensive | Expensive | Low to Moderate |

In the present study, we report AFM based nanoscale friction analysis and evolution of ordered graphene layers over the *a*-C tribofilms in the wear track of two different sets of UNCD films grown under Ar and $N_2$ atmospheres. These UNCD films were pre-studied with various macroscale measurements such as Raman spectroscopy, HRTEM, EELS and XPS [6]. Herein, our FFM study provides a direct micro-/nanoscopic evidence for the formation of 2D graphene layers through shear induced graphitization and these secondary phases have distinct friction behavior compared to the *a*-C tribofilm matrix. Moreover, these layered graphene structures play a decisive role on the ultralow friction and wear in these UNCD films.

## 2. Experimental

Two sets of UNCD films grown under Ar and $N_2$ atmospheres (labeled as $UNCD_{Ar}$ and $UNCD_N$, respectively) are considered for the study. The physical and macroscopic tribological properties of the films are reported elsewhere [6]. Herein, in order to understand the origin of ultralow friction and wear, a systematic nanoscale friction analysis is performed using FFM inside the wear track that were run for different sliding distances, viz. 30, 100 and 500 m for $UNCD_{Ar}$ and 2.5, 100, and 500 m for $UNCD_N$ films. Topography and friction force images are acquired simultaneously using AFM ( NT-MDT, Russia) with soft Si cantilevers of ~ 350 μm length and ~ 0.05 N/m spring constant. AFM measurements are repeated with typical scan area of ~ 40 x 100 μm² over a section across the wear track whose width is about 255 and 310 μm for $UNCD_N$ and $UNCD_{Ar}$ films, respectively. Further, FFM measurements are repeated on several locations in the wear track encompassing FLG nanostructures at different AFM tip normal load. In addition, the local elastic modulus of unworn and worn out surfaces are estimated by measuring contact resonance frequency (CRF) using stiff Si cantilevers in atomic force acoustic microscopy (AFAM). The experimental details of AFAM and estimation of elastic modulus can be found elsewhere [46–48]. A typical normal load of ~ 1 nN and 1 µN is applied on AFM tip for FFM and AFAM measurements respectively, unless otherwise specifically mentioned in the text.

## 3. Results

Before going into the FFM study, a brief summary of previous results is valuable. The macroscale tribometric studies reveal that the $UNCD_{Ar}$ films exhibit an ultrahigh wear resistance with average saturated CoF of ~ 0.08. Further, it is observed that the CoF is higher ( ~ 0.27 ) during the initial stage of sliding and it gradually decreases to lower value (0.08) after a long run-in distance of about 100 m. On the other hand, $UNCD_N$ films have ultralow friction with saturated CoF of ~ 0.04 with negligible run-in distance, however, the wear rate is relatively higher compared to $UNCD_{Ar}$ films. Also, the hardness (elastic modulus) of pristine $UNCD_{Ar}$ and $UNCD_N$ films is found to be ~ 23 (215 GPa) and 14.8 (165 GPa) GPa respectively, as measured by nanoindentation. Based on Raman spectroscopy, HRTEM, EELS and XPS analysis, the pristine $UNCD_{Ar}$ films were found to have higher $sp^3/sp^2$ ratio with thin *a*-C layer at grain boundaries (GBs) while diamond grains are encapsulated with thick nanographitic layers at GBs of pristine $UNCD_N$ films [6]. In addition, the post wear analysis of the tribo tracks by the above mentioned techniques clearly revealed the presence of tribofilms with excess amount of graphitic phase [6] but they could not provide a conclusive remark for the difference in the tribological properties of these UNCD films.

## 3.1 Frictional force microscopy

Figures 1a (1c) ((1e)) and 1b (1d) ((1f)) present the simultaneously acquired topography and friction force images respectively, on wear track of $UNCD_{Ar}$ film that undergone macroscopic wear test for sliding distance of 30 m (100 m) ((500 m)). Note that the bright and dark areas in the

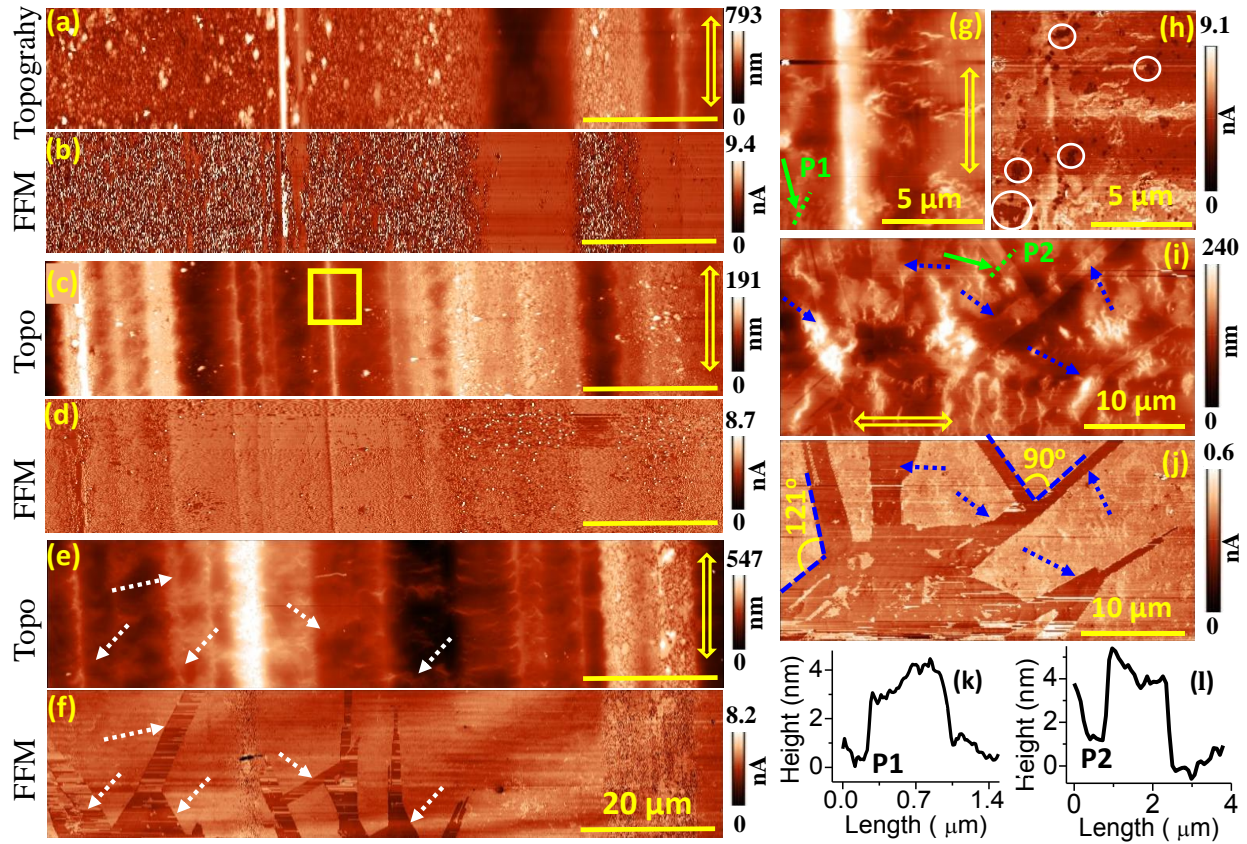

Fig. 1. Simultaneously acquired AFM topography (a,c,e) and frictional force (b,d,f) images of the wear track in the UNCD$_{Ar}$ that undergone wear test for the sliding distance of 30, 100 and 500 m, respectively. A magnified part of the topography (g) and friction (h) of wear track (100 m sliding distance, marked in image 1c). The selected area topography (i) and friction (j) of the wear track that ran for 500 m sliding distance. The arrow marks in Fig. 1(e,g,i) represent the presence of FLG structures which are not obviously visible due to a large height variation in the matrix. However, the presence of FLG can be obviously noticed in the corresponding friction images of Fig. 1(f,h,j) with much lower friction values. The height profiles across a FLG structures marked as P1 and P2 in Fig.1g and 1i are given in Fig. 1k and 1*l*, respectively. The double arrow marks in the images (a, c, e, g, i) indicate the sliding direction. Scale bar = 20 μm for the images in 1a-1f. The angle between the graphene edges are found to be ~ 90 and 120° as marked in the Fig.1j.

topography (FFM) represent the objects with maximum and minimum height (friction force), respectively. The wear track of UNCD$_{Ar}$ film is partially worn out after 30 m run-in distance, as shown in Fig. 1a and the wear scratches introduce very high root mean square (rms) roughness of 22.4 nm, measured over the entire image. The partially worn out surfaces have lower friction as compared to pristine surface, as can be seen in Fig. 1b. After a sliding distance of 100 m, the surface of wear track is slightly smoothened due to the movement of tribofilms (rms roughness = 16.9 nm) as depicted Fig.1c. The Fig. 1d shows the FFM image of the corresponding area in the wear track that displays both low and high friction areas. A careful observation of the magnified part of the topography ( Fig. 1g) reveals the presence of secondary phase, over the tribofilm matrix, as isolated bright batches with nearly flat top surface at an elevated height of (3.0 ± 1.0) nm and few hundreds of nanometer in lateral size in the wear track. Here, the error value is obtained from statistical mean of height measured on multiple particles. The corresponding FFM mapping (Fig. 1h) also displays dark regions (indicating lower friction force) wherever secondary phases are present in the tribofilms and a few such areas are marked as circles in Fig. 1h. The height profile of one such cluster labelled as P1 in Fig. 1g is shown in Fig. 1k and it has the height of ~ 2 nm with a flat top surface and sharp edges at the periphery.

Figures 1e, 1i and 1f, 1j are topography and friction images respectively, recorded at two different locations in wear track of UNCD$_{Ar}$ films that ran for 500 m sliding distance. The topographic images become more textured due to the wear ball induced scratches along the wear direction and the formation of tribofilms which move normal to the sliding direction, as can be evidenced in Fig.1e and 1i. The calculated rms roughness is found to be 24.7 and 7.2 nm for the wear track shown in Fig.1e and 1i, respectively. Further, Fig. 1i shows prominent and large networks of 2D nanoribbon like structures with height of ~ (3 ± 2.0) nm and length > 15 µm. The width of the nanoribbons varies from ~ 1 to 10 µm. A careful look at Fig. 1e also reveals the presence of similar 2D nanoribbon networks and they are indicated by arrow marks as guide to eye. The presence of such 2D nanoribbons make striking difference in corresponding FFM images with lowest friction values (dark regions) as illustrated by arrow marks in Figs. 1f and 1j. The Fig.1*l* shows the height profile of one such nanoribbon indicated by an arrow in Fig.1i at position P2. The height and width of the nanoribbons are 4 nm and 1.8 µm, respectively. In addition, these 2D nanoribbon structures are mostly oriented along the sliding direction and a few branches are noticed in normal to the sliding direction. Apart from the large networks of 2D nanoribbon structures, there are also smaller clusters with unique friction characteristics in the wear track that was formed with 500 m sliding distance. On the other hand, such secondary phase nanoclusters or 2D nanoribbon structures with unique low friction characteristics are not detected on the wear track of $UNCD_{Ar}$ films for the sliding distance of 30 m, even though $UNCD_{Ar}$ films had undergone mild wear at this sliding distance. We note here that the observation of 2D structures with atomic step height and flat top surface suggest the signature of FLG structures [36]. Moreover, the observation of low friction on the FLG structures reaffirms the presence of layered graphitic structure which is well known for low friction [36,39,43,49]. Here, the typical thickness of 2 nm indicates that the FLG structures have about 5 atomic layers of graphene since the monolayer graphene has the thickness of ~ 0.34 nm.

Figure 2a (2c) ((2e)) and 2b (2d) ((2f)) depict the topography and friction images respectively, in the wear track of $UNCD_N$ films that ran for 2.5 m (100 m) ((500 m)) sliding distances. The rms roughness of the wear track is about 27.9, 27.1 and 86.2 nm for sliding distance of 2.5, 100 and 500 m, respectively. The surface of the $UNCD_N$ films undergo smoothening even after the sliding distance of 2.5 m (Fig. 2a). With increasing the sliding distance to 100 m, the

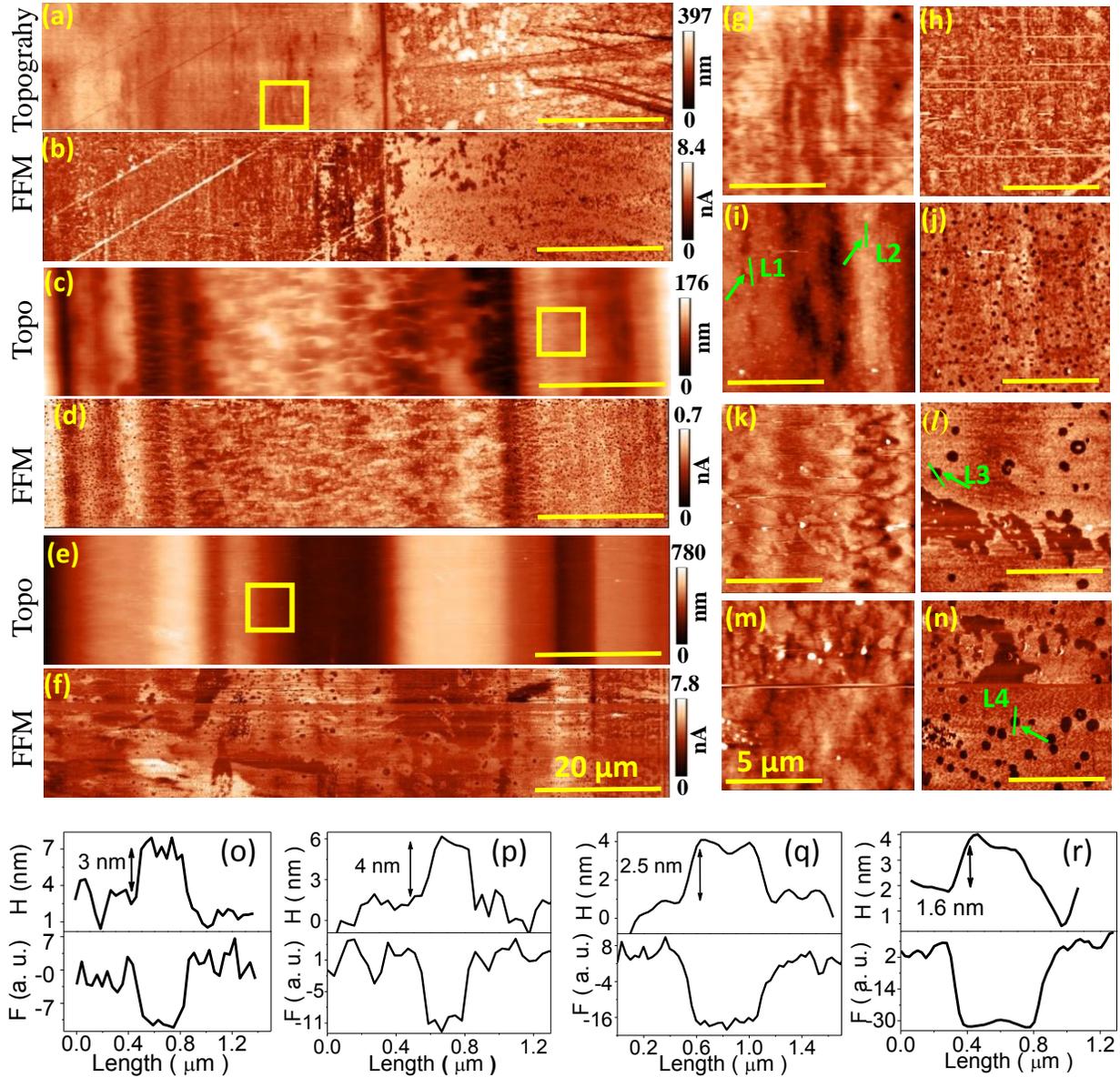

Fig. 2. AFM topographies (a,c,e) and corresponding friction (b,d,f) images of UNCD$_N$ films measured inside the wear track that ran for 2.5, 100, and 500 m sliding distances, respectively. A magnified part of the topography and friction images (marked as square box area on the topographic images in Fig 2 (a,c,e)) are given in the images Fig. 2(g,i,k) and Fig. 2 (h,j,*l*), respectively. (m) Topography and (n) friction mapping recorded at different location in the wear track that undergone 500 m sliding distance. The scale bar is 20 and 5 μm for the images 2(a-f) and 2(g-*l*), respectively. The plots given in Fig 2o and 2p (2q and 2r) represent the line profile of height and corresponding friction force measured across graphitic structures on the wear track of 100 m (500 m) siding distance, at the locations marked as L1, L2, L3 and L4.

topography becomes much rougher with movement of tribofilms normal to the sliding direction (Fig.2c). After 500 m sliding, the wear track appears smooth due to the rearrangement of tribofilms. However, the depth of trenches induced by the roughness of the wear ball is higher (Fig.2e) compared to that observed at shorter sliding distance. A magnified part of the topography and friction images of the wear tracks are shown in Fig. 2g-2*l*. There is no signature of FLG secondary phases in the magnified part of the topography (Fig.2g) and friction mapping (Fig.2h) of wear track after 2.5 m sliding distance. After 100 m sliding distance, the friction mapping (Fig. 2d and 2j) contains large number density of dark spots indicating lower friction areas in the wear track, that can be attributed to the presence of graphitic nanoclusters which are not obviously seen from topography (Fig. 2c). However, the magnified part of the topography confirms the presence of FLG nanoclusters with large number density in the wear track. Similarly a large amount of FLG nanoclusters are also present in the wear track that ran for 500 m sliding distance (Fig 2k, 2m).

Figures 2o and 2p (2q and 2r) show the line profile of height and friction force recorded on selected FLG structures marked as L1 and L2 ( L3 and L4) in the Fig.2i and Figs. 2*l* & 2n, respectively on the wear track of 100 m (500 m) siding distance. Moreover, based on height measurements, the average height and lateral size of the nanoclusters are ~ $(5.0 \pm 3.0)$ nm [$(4 \pm 2.5)$ nm] and $(250 \pm 50)$ nm [$(600 \pm 300)$ nm] for sliding distance of 100 m [500 m], respectively. These FLG nanostructures with uniform diameter are homogeneously distributed over entire region of the wear track after sliding for 100 m while the FLGs are inhomogeneously distributed over entire region of the track after 500 m sliding distance.

For nanoscale friction analysis of these FLG nanostructures, FFM measurements were performed on selected area by varying normal load. Sequence of friction images were recorded at constant normal load over a selected area and for each image, the normal load on AFM cantilever was increased in steps varying from 0.2 to 30 nN. Typical topography and friction images, recorded in the wear tracks of UNCD$_{Ar}$ (500 m) for normal load of 1 and 30 nN respectively, are shown in Figs. 3a, 3c and 3b, 3d. The measured height of 0.8 nm indicates the presence of bilayer graphene in the selected region (Fig. 3e, 3f). We also note here that the friction force signal was collected twice in each specified location viz. during the forward and reverse scan of AFM tip and the net friction force mapping was constructed by subtracting one from the other. Subsequently, the distribution of friction force in particular area is analyzed by constructing a histogram from each friction image and the typical histograms for the normal load of 1 and 30 nN are shown in Fig. 3i. Further, the mean statistical friction force corresponding to the low (FLG structures) and high friction ( matrix) regions are calculated from the histogram. These friction force values are plotted as a function of normal load, as shown in Fig. 3j and the slope of the plot gives the CoF. As shown in Fig. 3j, the friction force increases as a function of normal load for both FLG nanoclusters and the a-C tribofilm matrix. The increase in friction is attributed to the puckering effect which arises due to elastic deformation of the FLG structures with increase in normal load [36,39]. The weak Van der Waals force between the graphene layers allows puckering which results in reduction of lateral dimension of the FLG structures, as can be seen from Fig. 3d. Based on the slope of 10 and 27.7 (unnormalized), the relative CoF of FLG structure is about three times lower than the matrix in the track. A similar analysis is also performed for the wear track of $UNCD_N$ (500 m) and the plot is shown in Fig. 4. The height of the FLG structure is 4 nm, comprising 10 layers of graphene. In this wear track also the friction force increases with normal load. The slopes of the curves for FLG structures (low friction areas) and matrix (high friction areas) are found to be 3.1 and 5.4 respectively. As similar to $UNCD_{Ar}$ films, the relative CoF is lower for FLG structures in $UNCD_N$ as compared to the matrix. In addition, the relative CoF of FLG structures and the matrix in $UNCD_N$ films is ~ 3 and 5 times lower than that of $UNCD_{Ar}$ films, respectively. These results are consistent with macroscopic tribo measurements which shows the average saturated CoF of ~ 0.08 and 0.04 for $UNCD_{Ar}$ and $UNCD_N$ films [6].

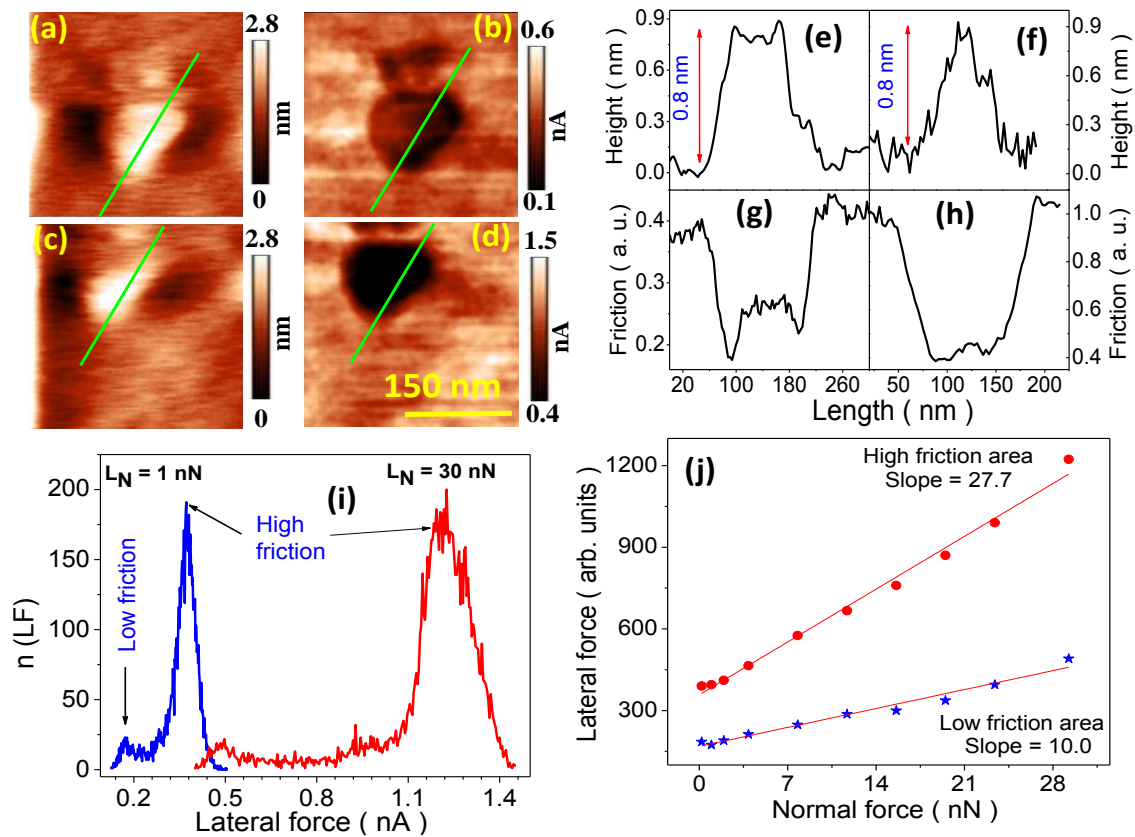

Fig. 3. (a,c) Topography and (b,d) friction images on the wear track of UNCD$_{Ar}$ measured at AFM tip normal load of (a,b) 1 and (c,d) 30 nN, respectively. (e-h) Line profiles of height and friction force measured on respective images of 3(a-d). (i) Histograms constructed from the friction force images given in Fig. 3 b (the spectra in red color) and 3d (the spectra in blue color). (j) The variation of friction force as a function of normal load on UNCD$_{Ar}$ film; the two slopes indicate the difference in coefficient of friction in the low and high friction areas.

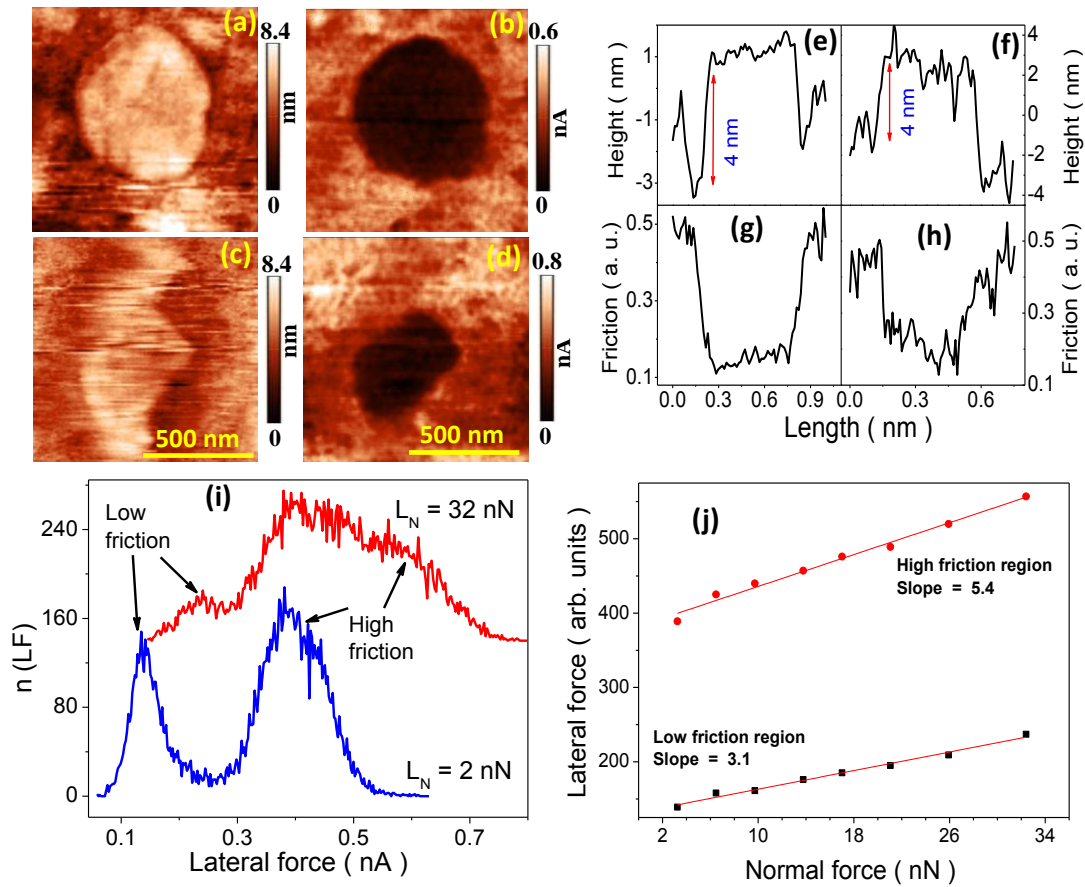

Fig. 4. (a,c) Topography and (b,d) friction images on the wear track of UNCD$_N$ measured at AFM tip normal load of (a,b) 1 and (c,d) 30 nN, respectively. (e-h) Line profiles of height and friction force measured on respective images of 4(a-d). (i) Histograms constructed from the friction force images given in Fig. 4b (the spectra in red color) and 4d (the spectra in blue color). (j) The variation of friction force as a function of normal load on UNCD$_N$ film

Further, the layered nature of the newly formed FLG structures and their binding on the tribofilms in the wear track are tested by measuring topography and friction mapping at slightly higher normal load on the cantilever. Fig. 5a and 5d show the simultaneously measured topography and friction force mapping on wear track of $UNCD_{Ar}$ (500 m) at a normal load of 4 nN on the cantilever. As discussed earlier, the FLG structures have lower friction (darker area) as compared to the tribofilm of the matrix. Also, some part of the FLG structures (arrow mark in Fig. 5a) have already started peeling off and it is also reflected as bright areas in the FFM mapping (arrow mark in Fig. 5d). Subsequent imaging at higher load of 10 nN on the same location reveals that the FLG structures are completely removed from the matrix as can be seen from Fig. 5b. It is also noticed that the previously known low friction area (darker regions) switches into higher friction area (bright regions in Fig. 5e) immediately after the removal of FLGs. Moreover, the bright areas (high friction regions) on right side ligament of the Y junction (Fig. 5d) indicate that the partial FLG structures are peeled off during the forward scan, which is not reflected in the topography (Fig. 5a), but it shows the effect on friction force image. In order to verify the flipping of friction from low to high upon removal of FLG structures, an additional FFM measurement at 2 nN normal load was performed on larger area encompassing the previously measured area, and the topography and FFM image are shown in Fig. 5c and 5f, respectively. This measurement confirms the lower friction of FLG when the physical structure is intact (upper part of Fig 5c). Subsequently, after the FLG structures are physically removed, the underneath area becomes even higher in friction than the matrix. Since the friction measurements were performed immediately after removal of FLGs, the newly exposed area exhibits high friction due to inherent nature of the tribofilms which consist of highly disordered *a*-C with a large amount of dangling bonds that increase the friction. In contrast, the smaller dimension FLG structures that are present on the wear track are not being completely removed, as compared to large networks of FLG. However, at higher normal load on AFM tip, the smaller FLG structures undergo mechanical deformation as can be seen in Fig. 3c and Fig. 4c.

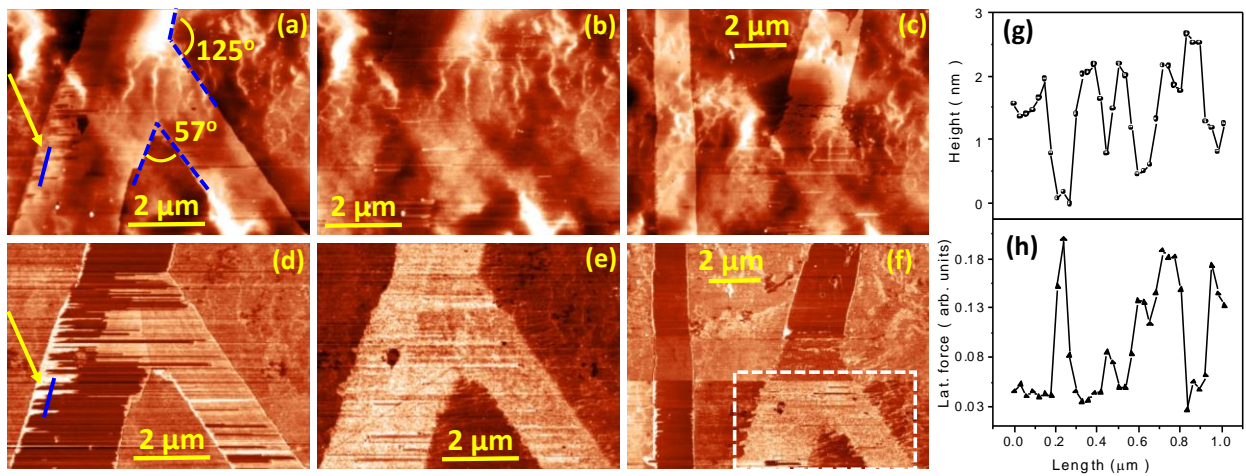

Fig. 5. Simultaneously acquired topography and friction mapping on a few layer graphene structures. Mappings are recorded at normal load of 2 nN for the images given in 5a,5c,5d, and 5f, and at 10 nN for the images in 5b an 5e. The rectangle symbol on Fig. 5f indicates the previously AFM measured area at higher load of 10 nN. The line profile of (g) height and (h) lateral force measured along the line marked in the image 5(a) and 5(d), respectively.

Further, the adhesion behavior of the newly formed 2D structures are measured on wear track of $UNCD_{Ar}$ film by varying the tip scan velocity under constant normal loads of 1 and 8 nN. Fig. 6a and 6b depict the topography and friction mapping respectively, measured in the wear track at normal load of 1 nN. Here, the friction is lower on the 2D structures as compared to matrix a-C tribolayer. Further, sequence of friction images were recorded at constant normal load over a selected area and for each image, the AFM tip scan velocity is varied from 0.5 to 5 μm/s. Then, histogram is constructed for each friction image to extract the average lateral force on the low and friction areas for different scan velocities. Fig. 6c and 6d show the variation of lateral force as a function of scan velocity at normal load of 1 and 8 nN, respectively. As can be seen from Fig. 6c & 6d, the low and high friction areas display negative and positive slope for friction as a function of scan velocity. The negative slope of friction in the FLG nanocluster is attributed to meniscus force which arises due to the condensation water molecules between tip and sample surface. The slower scan rate increases the contact area leading to higher friction and also, this behavior indicates hydrophilic nature of the FLG nanoclusters. On the other hand, the positive friction is associated with the hydrophobic character of a-C matrix which is similar to earlier report [50]. Though the friction measurements could not be performed under higher scan velocity due to structural deformation of FLG, this observation clearly reveals the different surface chemistry of FLG structures and a-C matrix. In addition, the elastic properties also influence the tribological properties and hence, the elastic properties of these UNCD films and tribolayers are measured using AFAM and are discussed below.

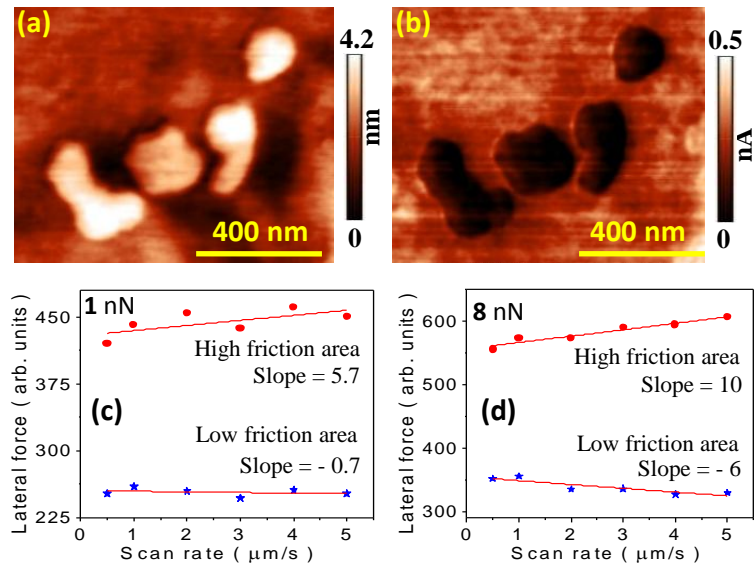

Fig. 6. Simultaneously acquired topography (a) and friction (b) mapping on a few layer graphene structures, measured at normal load of 1 nN. The height of the individual structure is ~ 1.6, 0.8, 2.4 and 2.0 nm respectively, from left to right. The variation of lateral force as a function of AFM tip scan velocity at (c) 1 nN and (d) 8 nN normal loads indicating the opposite behavior of lateral force on low and high friction areas in the wear track.

**3.2 Atomic force acoustic microscopy**

Atomic force acoustic microscopy is a versatile technique to probe local elastic properties of materials at nanometer-scale. In AFAM measurements, the sample is mounted on a piezoelectric transducer which generates ultrasound frequencies upto 5 MHz. When the sample is vibrated at a drive frequency, the AFM cantilever, in contact with sample, is also forced to vibrate at contact resonance frequency (CRF) which mainly depends on the elastic modulus of the tip and sample, and contact load on the AFM tip. By analyzing CRF, the elastic modulus of the samples can be calculated using the equation (1) when the cantilever properties are known [46].

$$E_s^* = E_{ref}^* \left(\frac{k_s^*}{k_{ref}^*}\right)^n \quad \text{............... (1)}$$

where, $E_s$ and $E_{ref}$ are the elastic modulus of unknown and reference sample, respectively. The $k_s^*$ and $k_{ref}^*$ are the effective contact stiffness parameter of unknown and reference samples, respectively; Also, $k^* = f_{CRF}^2 - f_0^2$, depends on the contact resonance frequency of tip-sample in collective vibration ($f_{CRF}$) and free resonance frequency ($f_0$) of the cantilever in air. Fig. 7 shows the CRF spectra of the pristine UNCD$_{Ar}$, UNCD$_N$ films and Si (100) wafer (reference). Based on equation 1 and $E_{ref}$ of 170 GPa for Si (100), the elastic modulus of the $UNCD_{Ar}$, and $UNCD_N$ films is estimated to be $184 \pm 10$ and $163 \pm 10$ GPa, respectively. The estimated elastic modulus of the films is almost close to the value of 215 and 165 GPa respectively, measured by nanoindentation [6]. In order to have better understanding on the local elastic modulus of the tribofilms, the matrix and FLG structures in the wear track, AFAM CRF mapping was employed and the results are discussed below.

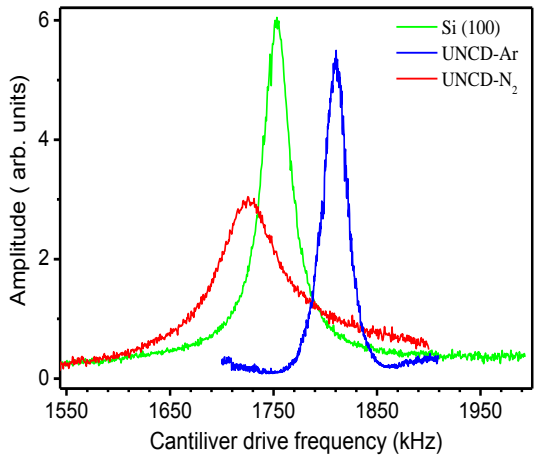

Fig. 7. The contact resonant frequency spectra measured on the pristine surface of UNCD$_{Ar}$ and UNCD$_N$ films. The CRF spectrum on Si (100) is also given for reference.

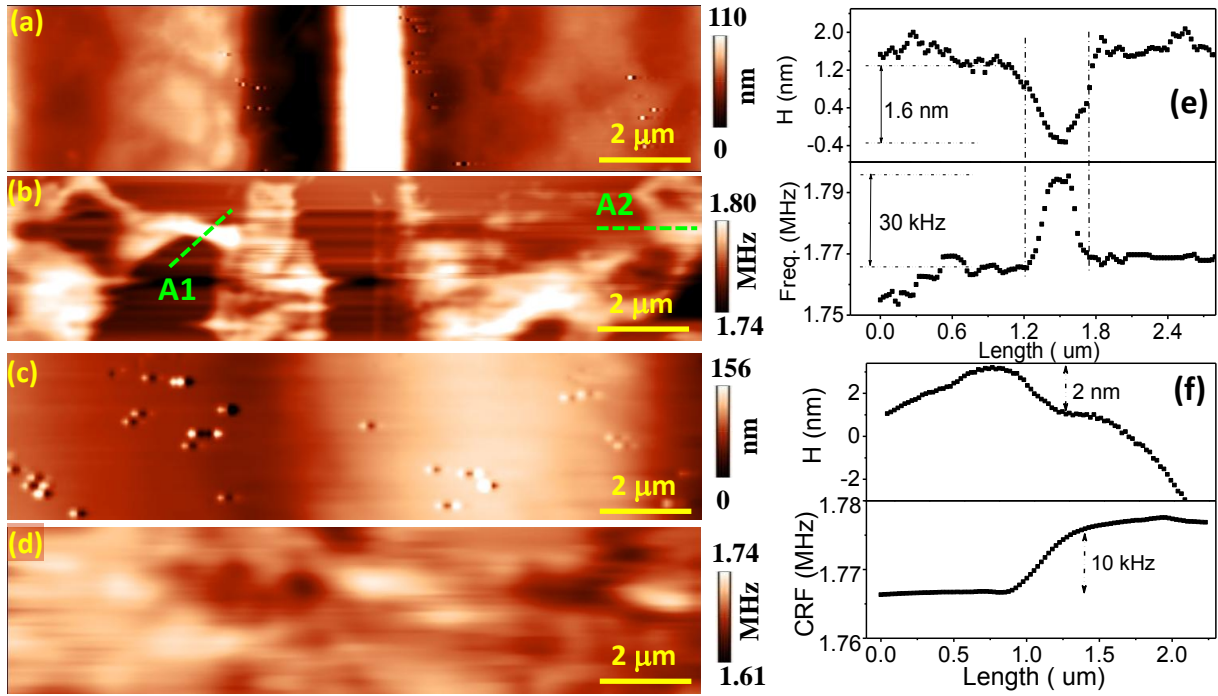

Fig. 8. Simultaneously acquired topography (a,c) and contact resonance frequency (b,d) mapping on the wear track of $UNCD_{Ar}$ (a,b) and $UNCD_N$ (c,d) films. (e,f) The line profiles of height and contact resonance frequency measured across graphitic structures on positions A1 and A2.

Figures 8a and 8b (8c and 8d) display the simultaneously acquired topography and CRF mapping respectively, on the wear track of $UNCD_{Ar}$ ($UNCD_N$) film with 500 m sliding distance.

Similar to friction mapping, CRF mapping (Fig. 8b) also displays both bright and dark areas corresponding to high and low elastic modulus, respectively. A careful observation of the topography corresponding to the brighter regions indicates the presence of atomically thin FLG structures which are well known for their lower elastic modulus as compared to other carbon allotropes. The line profiles of height and CRF variation across a FLG structure (marked as A1 and A2 in Fig. 8b) are given in Fig. 8e and 8f, respectively. The CRF decreases down to 30 kHz (~ 2 %) on FLG structures ( ~ 2 nm thickness, as shown in Fig. 8e) compared to the tribofilm in the wear track. Note that the observed lower CRF indicates reduction in effective elastic modulus of the coupled FLG structures and the *a*-C tribofilm matrix. Similar AFAM measurements on $UNCD_N$ reveal that only small clusters having low elastic modulus are present inside the wear track. The estimated elastic modulus of the tribofilms in the wear track of $UNCD_{Ar}$ is larger than that of the $UNCD_N$ films due to higher $sp^3$ content and the inherent mechanical characteristics of the unworn UNCD films. Moreover, a significant amount of mixed low and high elastic modulus regions are present in the wear track of $UNCD_{Ar}$ films while a homogeneous distribution of elastic modulus regions with isolated low elastic modulus are observed in the wear track of $UNCD_N$ films. Thus, the AFAM studies reveal the variation in elastic modulus of the tribofilms and corroborate the FFM studies in terms of distribution of FLG structures in the wear tracks of both the UNCD films.

## 4. Discussion

Based on the experimental observation, a brief summary of the results are discussed here. The surface topography and friction force mapping unveil that the wear track of $UNCD_{Ar}$ films consists of a large network of 2D FLG nanoribbons with thickness of ~ (3 ± 2) nm, width and length of several micrometers. On the other hand, the wear tracks of $UNCD_N$ films only have small

FLG structures with a large number density, thickness of ~ (4 ± 2.5) nm and lateral size of < 1 μm. Besides, we observe a mixture of bilayer, trilayer and multilayer graphene nanosheets with < 15 layers on the wear tracks of both these UNCD films as confirmed by the atomic step height measurements by AFM. Nevertheless, the statistical distribution of height analysis suggest that the FLG structures in the wear track of $UNCD_{Ar}$ films have lower thickness than that of $UNCD_N$ films. Further, the local friction is much lower in specific areas in the wear track wherever the FLG structures are present. Furthermore, the relative nanoscale CoF of the tribofilms on the wear tracks of $UNCD_N$ is lower than that of the tribofilms in $UNCD_{Ar}$, as discussed in Fig. 3 and Fig. 4. This result supports the macroscopic tribo measurements which gives the average saturated CoF of 0.08 and 0.04 for $UNCD_{Ar}$ and $UNCD_N$ films, respectively. The lower CoF in $UNCD_N$ films can be attributed to the presence of thicker FLG structures than in $UNCD_{Ar}$ films since the CoF is thickness dependent and it is lowest for bulk graphite and increases with decrease in layer thickness [38].

Another important observation is that the local friction in the tribofilms increases when the protective FLG structures are removed. This observation endorses that the matrix, the dangling bonds in the *a*-C tribofilms, undergoes surface passivation by termination of –H and –OH functional groups originated from the dissociation of $H_2O$ molecules in the ambient atmosphere because of the tribochemical reactions at the sliding interface [5,18,51,52]. Especially, the electrostatic repulsion offered by H- termination in diamond or DLC can lead to the superlubricity between the counterfaces [52]. Further, the C-H bonding in $sp^3$-rich matrix is more stable than the $sp^2$-rich matrix due to the high binding energy. Moreover, the presence of higher $sp^3$ content in tribofilms of $UNCD_{Ar}$ films is also reflected as higher elastic modulus than that of the $UNCD_N$ films, as estimated by AFAM. Raman spectroscopic studies in the wear track clearly confirm that UNCD$_{Ar}$ has higher *sp*$^3$ content compared to UNCD$_N$ films [6]. Hence, perhaps, the higher *sp*$^3$ C-C bonding in the tribofilms of UNCD$_{Ar}$ films had also facilitated a more efficient surface passivation that can result in ultralow friction with a minimum wear rate. In contrast, the higher *sp*$^2$ C=C bonding in the tribofilms of UNCD$_N$ films enables the formation of C-OH, C-O or C=O bonding that are highly unstable leading to CO or CO$_2$ molecules and make the surface to wear out at higher rate. Moreover, it has been recently reported that the in-plane tensile fracture strength and critical compressive strain to failure decrease with increase in layer thickness of graphitic structures [53,54]. Thus, the FLG structures with slightly high thickness in UNCD$_N$ films undergo failure at higher rate than that of tribofilms in UNCD$_{Ar}$ films and this makes the wear rate to be higher in UNCD$_N$.

Based on the above discussion, we propose here that large networks of FLG structures act as lubricant as well as protecting layer that results in ultralow friction and high wear resistance in UNCD$_{Ar}$ films [7,8,32]. Also, Wijk et al [8] predicted that the FLG act as wear protection of diamond, by preventing further plastic deformation of diamond into *a*-C. On the other hand, the UNCD$_N$ films have large number density of FLG nanoclusters which act as efficient lubricant resulting in ultralow friction. However, the high wear rate in UNCD$_N$ films can be associated with inability to form larger networks of FLG nanostructures. Also, the adhesion of newly formed layered FLG structures over the a-C matrix tribofilm is poor. This results in the removal of tribofilms in the wear track leading to a significant wear rate on both the films. However, the higher wear rate in UNCD$_N$ films than that of the UNCD$_{Ar}$ films, can be associated with the tribofilms with higher *sp*$^2$ content which are easily moved out of the track due to the roughness of the counter body [3,6].

Despite the above discussion, the question remains that whether the newly formed low friction structures are few layer graphene or not ? or, is it $sp^2$-rich a-C adlayer since there is no alternative direct evidence. Here, we reiterate the following evidences revealed by combined AFM and FFM measurements.

1. Atomic step height and low friction behavior of FLG structures as shown in Figs. 1- 6.
2. Partial cleavage with sharp edges and variation of friction with layer thickness ( Fig. 5)
3. Angle between adjacent edges are about 60, 90 and 120º as indicated in Figs. 1 and 5. These angles are characteristic of graphene edge chirality that arises due to armchair and zigzag edges [55].
4. Friction force decreases with increase in AFM tip scan velocity on FLG structures while an opposite behavior is observed for a-C matrix (Fig. 6)

All the above mentioned four points are characteristics of layered graphene structures. Suppose if we were to assume the newly formed structure as $sp^2$-rich a-C, it is impossible to cleave the partial top layer with sharp edges and also, the friction does not change with thickness. Moreover, the adhesion / surface chemistry cannot be opposite to the a-C matrix. Based on these facts, we infer that the newly formed tribolayers are indeed layered graphene structures. However, a direct measurement like TERS can confirm the layered structure with distinct Raman characteristics of FLG structures. Also, HRTEM, scanning tunneling microscopy and Kelvin probe microscopy can shed more light on the layered nature of the FLG. Nevertheless, these techniques do also have several drawbacks to study of lateral tribolayers which are sparsely and randomly present on a rough wear track. Yet, AFM is a powerful and relatively simple tool which helps to understand the complex wear track with the support of well-established characteristics of carbon materials.

Finally, it is very intuitive to know the driving force for the nucleation of ordered graphene structures over the tribolayers. Here, the rms roughness and wear test environments are almost identical for both the films, though, the structural characteristics, $sp^3/sp^2$ ratio, and hardness are different and they might play a significant role on wear and friction. It is mostly accepted that the shear force induces $sp^3$-$sp^2$ order – disorder transformation which leads to the formation of $sp^2$-rich $a$-C tribolayers at the sliding interface [19]. Further, the shear localization mechanism suggests that rehybridization occurs at the tribo-interface with structural phase transformation, covalent bond reorientation and local structural ordering [28]. Thus, the clusters of ordered crystallites with ~ 1 nm in size nucleate in the $a$-C matrix which has already been experimentally evidenced by HRTEM [2,6,10,35]. However, it was not clear so far that whether these small crystallites grow into large size graphene layers or not. In fact, this work by AFM studies provide evidences of the growth of micro-/nanoclusters with ordered graphene layers or a large network of graphene nanoribbons in the wear track. Moreover, it should be noted here that the shear force induced energy dissipation at the sliding interface is very high for $UNCD_{Ar}$ films due to the initial high CoF along with long run-in distance, high hardness and elastic modulus which are leading to the formation of large networked FLG structures. However, the energy dissipation is much lower for $UNCD_N$ films due to the higher $sp^2$ content which makes negligible run-in distance and lower mechanical properties. Thus, the structural and mechanical properties play a significant role on the tribological properties of UNCD films. Also, we cannot rule out the role of H - surface passivation of the tribofilms which can play a significant role on friction and wear in UNCD films.

## 5. Conclusions

The formation of bilayer, trilayer, and a few layer graphene-like structures which arise due to shear induced graphitization are observed over the amorphous carbon ($a$-C) tribofilms on two different sets of UNCD films using frictional force microscopy. Since the newly evolved few layer graphene structures act as lubricant as well as wear protector, UNCD films with large networked few layer graphene structures exhibit high wear resistance with low friction. In contrast, UNCD films only with isolated FLG structures display ultralow friction but it could not protect wear. However, the growth of ordered graphene structures over the tribolayers depends on the structural and mechanical characteristics of the UNCD films. In nutshell, the shear induced graphitization and the subsequent formation of ordered graphene layers play a major role on the ultralow friction and wear in UNCD films. Further, we cannot rule out the effect of H- surface passivation of the $sp^2$-rich $a$-C tribofilms on friction and wear. Carefully designed experiments are required to elucidate the driving force behind the formation of a large networks of ordered graphene structures since the tribochemical reactions are complex. Moreover, despite the 2D materials have excellent frictional properties, they are easily affected by surface conditions, resulting in poor tribological results. Intensive study of atomic level friction using emerging 2D materials is need of the hour in the field of tribological research. Overall, the frictional force microcopy, combined with other macro-scale measurements, can serve as an excellent tool for understanding the tribology of carbon materials.

## Acknowledgements

Authors acknowledge Dr. Niranjan Kumar, erstwhile member of SND for originating the work and fruitful discussion on earlier studies with the same samples. We also thank Prof. I-Nan Lin, Tamkang University, Taiwan for providing the UNCD films and discussion. One of the authors, K.G., acknowledges Dr. Shaju K. Albert, MSG, IGCAR for his constant support and encouragement.

## References

[1] X. Chen, J. Li, Superlubricity of carbon nanostructures, Carbon N. Y. 158 (2020) 1–23. https://doi.org/10.1016/j.carbon.2019.11.077.

[2] R.A. Bernal, R.W. Carpick, Visualization of nanoscale wear mechanisms in ultrananocrystalline diamond by in-situ TEM tribometry, Carbon N. Y. 154 (2019) 132–139. https://doi.org/10.1016/j.carbon.2019.07.082.

[3] P.K. Ajikumar, K. Ganesan, N. Kumar, T.R. Ravindran, S. Kalavathi, M. Kamruddin, Role of microstructure and structural disorder on tribological properties of polycrystalline diamond films, Appl. Surf. Sci. 469 (2019) 10–17. https://doi.org/10.1016/j.apsusc.2018.10.265.

[4] R. Rani, K. Panda, N. Kumar, K.J. Sankaran, K. Ganesan, I.N. Lin, Tribological Properties of Ultrananocrystalline Diamond Films in Inert and Reactive Tribo-Atmospheres: XPS Depth-Resolved Chemical Analysis, J. Phys. Chem. C. 122 (2018) 8602–8613. https://doi.org/10.1021/acs.jpcc.8b00856.

[5] A.R. Konicek, D.S. Grierson, P.U.P.A. Gilbert, W.G. Sawyer, A. V. Sumant, R.W. Carpick, Origin of ultralow friction and wear in ultrananocrystalline diamond, Phys. Rev. Lett. 100 (2008). https://doi.org/10.1103/PhysRevLett.100.235502.

[6] R. Rani, K.J. Sankaran, K. Panda, N. Kumar, K. Ganesan, S. Chakravarty, I.N. Lin, Tribofilm formation in ultrananocrystalline diamond film, Diam. Relat. Mater. 78 (2017) 12–23. https://doi.org/10.1016/j.diamond.2017.07.009.

[7] D. Berman, S.A. Deshmukh, S.K.R.S. Sankaranarayanan, A. Erdemir, A. V. Sumant, Macroscale superlubricity enabled by graphene nanoscroll formation, Science (80-. ). 348 (2015) 1118–1122. https://doi.org/10.1126/science.1262024.

[8] M.M. Van Wijk, A. Fasolino, Minimal graphene thickness for wear protection of diamond, AIP Adv. 5 (2015) 1–7. https://doi.org/10.1063/1.4905942.

[9] T. Kuwahara, G. Moras, M. Moseler, Friction Regimes of Water-Lubricated Diamond

(111): Role of Interfacial Ether Groups and Tribo-Induced Aromatic Surface Reconstructions, Phys. Rev. Lett. 119 (2017) 1–6. https://doi.org/10.1103/PhysRevLett.119.096101.

[10]　P. Manimunda, A. Al-Azizi, S.H. Kim, R.R. Chromik, Shear-Induced Structural Changes and Origin of Ultralow Friction of Hydrogenated Diamond-like Carbon (DLC) in Dry Environment, ACS Appl. Mater. Interfaces. 9 (2017) 16704–16714. https://doi.org/10.1021/acsami.7b03360.

[11]　J.M. Martin, A. Erdemir, Superlubricity: Friction's vanishing act, Phys. Today. 71 (2018) 40–45. https://doi.org/10.1063/PT.3.3897.

[12]　A. Erdemir, J.M. Martin, Superior wear resistance of diamond and DLC coatings, Curr. Opin. Solid State Mater. Sci. 22 (2018) 243–254. https://doi.org/10.1016/j.cossms.2018.11.003.

[13]　X. Yin, J. Zhang, T. Luo, B. Cao, J. Xu, X. Chen, J. Luo, Tribochemical mechanism of superlubricity in graphene quantum dots modified DLC films under high contact pressure, Carbon N. Y. 173 (2021) 329–338. https://doi.org/10.1016/j.carbon.2020.11.034.

[14]　Q. Yu, X. Chen, C. Zhang, J. Luo, Influence Factors on Mechanisms of Superlubricity in DLC Films: A Review, Front. Mech. Eng. 6 (2020) 1–17. https://doi.org/10.3389/fmech.2020.00065.

[15]　Z. Chen, X. He, C. Xiao, S.H. Kim, Effect of humidity on friction and Wear-A critical review, Lubricants. 6 (2018) 1–26. https://doi.org/10.3390/lubricants6030074.

[16]　E.E. Ashkinazi, V.S. Sedov, M.I. Petrzhik, D.N. Sovyk, A.A. Khomich, V.G. Ralchenko, D. V. Vinogradov, P.A. Tsygankov, I.N. Ushakova, A. V. Khomich, Effect of crystal structure on the tribological properties of diamond coatings on hard-alloy cutting tools, J. Frict. Wear. 38 (2017) 252–258. https://doi.org/10.3103/S1068366617030047.

[17]　M. Shabani, A. C.S., G. J.R., R.F. Silva, O. F.J., Effect of relative humidity and temperature on the tribology of multilayer micro/nanocrystalline CVD diamond coatings, Diam. Relat. Mater. 73 (2017) 190–198. https://doi.org/10.1016/j.diamond.2016.09.016.

[18] F. Mangolini, K.D. Koshigan, M.H. Van Benthem, J.A. Ohlhausen, J.B. McClimon, J. Hilbert, J. Fontaine, R.W. Carpick, How Hydrogen and Oxygen Vapor Affect the Tribochemistry of Silicon- and Oxygen-Containing Hydrogenated Amorphous Carbon: A Study Combining X-Ray Absorption Spectromicroscopy and Data Science Methods, Submitted. (2020). https://doi.org/10.1021/acsami.1c00090.

[19] L. Pastewka, S. Moser, P. Gumbsch, M. Moseler, Anisotropic mechanical amorphization drives wear in diamond, Nat. Mater. 10 (2011) 34–38. https://doi.org/10.1038/nmat2902.

[20] K. Panda, R. Rani, N. Kumar, K.J. Sankaran, J.Y. Park, K. Ganesan, I.N. Lin, Dynamic friction behavior of ultrananocrystalline diamond films: A depth-resolved chemical phase analysis, Ceram. Int. 45 (2019) 23418–23422. https://doi.org/10.1016/j.ceramint.2019.08.045.

[21] A.R. Konicek, D.S. Grierson, A. V. Sumant, T.A. Friedmann, J.P. Sullivan, P.U.P.A. Gilbert, W.G. Sawyer, R.W. Carpick, Influence of surface passivation on the friction and wear behavior of ultrananocrystalline diamond and tetrahedral amorphous carbon thin films, Phys. Rev. B - Condens. Matter Mater. Phys. 85 (2012) 1–13. https://doi.org/10.1103/PhysRevB.85.155448.

[22] R. Rani, K. Panda, N. Kumar, A.T. Kozakov, V.I. Kolesnikov, A.V. Sidashov, I.N. Lin, Tribological Properties of Ultrananocrystalline Diamond Films: Mechanochemical Transformation of Sliding Interfaces, Sci. Rep. 8 (2018). https://doi.org/10.1038/s41598-017-18425-4.

[23] Y. Wang, J. Guo, J. Zhang, Y. Qin, Ultralow friction regime from the in situ production of a richer fullerene-like nanostructured carbon in sliding contact, RSC Adv. 5 (2015) 106476–106484. https://doi.org/10.1039/c5ra20892k.

[24] T. Kunze, M. Posselt, S. Gemming, G. Seifert, A.R. Konicek, R.W. Carpick, L. Pastewka, M. Moseler, Wear, plasticity, and rehybridization in tetrahedral amorphous carbon, Tribol. Lett. 53 (2014) 119–126. https://doi.org/10.1007/s11249-013-0250-7.

[25] Y. Liu, L. Chen, B. Jiang, Y. Liu, B. Zhang, C. Xiao, J. Zhang, L. Qian, Origin of low

friction in hydrogenated diamond-like carbon films due to graphene nanoscroll formation depending on sliding mode: Unidirection and reciprocation, Carbon N. Y. 173 (2021) 696–704. https://doi.org/10.1016/j.carbon.2020.11.039.

[26] D. Berman, B. Narayanan, M.J. Cherukara, S.K.R.S. Sankaranarayanan, A. Erdemir, A. Zinovev, A. V. Sumant, Operando tribochemical formation of onion-like-carbon leads to macroscale superlubricity, Nat. Commun. 9 (2018). https://doi.org/10.1038/s41467-018-03549-6.

[27] X. Chen, C. Zhang, T. Kato, X.A. Yang, S. Wu, R. Wang, M. Nosaka, J. Luo, Evolution of tribo-induced interfacial nanostructures governing superlubricity in a-C:H and a-C:H:Si films, Nat. Commun. 8 (2017). https://doi.org/10.1038/s41467-017-01717-8.

[28] T.B. Ma, L.F. Wang, Y.Z. Hu, X. Li, H. Wang, A shear localization mechanism for lubricity of amorphous carbon materials, Sci. Rep. 4 (2014) 1–6. https://doi.org/10.1038/srep03662.

[29] L. Pastewka, S. Moser, M. Moseler, Atomistic insights into the running-in, lubrication, and failure of hydrogenated diamond-like carbon coatings, Tribol. Lett. 39 (2010) 49–61. https://doi.org/10.1007/s11249-009-9566-8.

[30] S. Kajita, M.C. Righi, A fundamental mechanism for carbon-film lubricity identified by means of ab initio molecular dynamics, Carbon N. Y. 103 (2016) 193–199. https://doi.org/10.1016/j.carbon.2016.02.078.

[31] S. Bhowmick, A. Banerji, A.T. Alpas, Friction reduction mechanisms in multilayer graphene sliding against hydrogenated diamond-like carbon, Carbon N. Y. 109 (2016) 795–804. https://doi.org/10.1016/j.carbon.2016.08.036.

[32] X. Yin, F. Wu, X. Chen, J. Xu, P. Wu, J. Li, C. Zhang, J. Luo, Graphene-induced reconstruction of the sliding interface assisting the improved lubricity of various tribo-couples, Mater. Des. 191 (2020) 1–9. https://doi.org/10.1016/j.matdes.2020.108661.

[33] A. Klemenz, L. Pastewka, S.G. Balakrishna, A. Caron, R. Bennewitz, M. Moseler, Atomic scale mechanisms of friction reduction and wear protection by graphene, Nano Lett. 14

(2014) 7145–7152. https://doi.org/10.1021/nl5037403.

[34] X. Li, A. Wang, K.R. Lee, Fundamental understanding on low-friction mechanisms at amorphous carbon interface from reactive molecular dynamics simulation, Carbon N. Y. 170 (2020) 621–629. https://doi.org/10.1016/j.carbon.2020.08.014.

[35] Y. Liu, E.I. Meletis, Evidence of graphitization of diamond-like carbon films during sliding wear, J. Mater. Sci. 32 (1997) 3491–3495. https://doi.org/10.1023/A:1018641304944.

[36] Y. Peng, Z. Wang, K. Zou, Friction and Wear Properties of Different Types of Graphene Nanosheets as Effective Solid Lubricants, Langmuir. 31 (2015) 7782–7791. https://doi.org/10.1021/acs.langmuir.5b00422.

[37] R. Mas-Ballesté, C. Gómez-Navarro, J. Gómez-Herrero, F. Zamora, 2D materials: To graphene and beyond, Nanoscale. 3 (2011) 20–30. https://doi.org/10.1039/c0nr00323a.

[38] C. Lee, Q. Li, W. Kalb, X.Z. Liu, H. Berger, R.W. Carpick, J. Hone, Frictional characteristics of atomically thin sheets, Science (80-. ). 328 (2010) 76–80. https://doi.org/10.1126/science.1184167.

[39] J.C. Spear, B.W. Ewers, J.D. Batteas, 2D-nanomaterials for controlling friction and wear at interfaces, Nano Today. 10 (2015) 301–314. https://doi.org/10.1016/j.nantod.2015.04.003.

[40] S. Zhang, T. Ma, A. Erdemir, Q. Li, Tribology of two-dimensional materials: From mechanisms to modulating strategies, Mater. Today. 26 (2019) 67–86. https://doi.org/10.1016/j.mattod.2018.12.002.

[41] J.H. Lee, D.H. Cho, B.H. Park, J.S. Choi, Nanotribology of 2D materials and their macroscopic applications, J. Phys. D. Appl. Phys. 53 (2020). https://doi.org/10.1088/1361-6463/ab9670.

[42] L.D.A. Prospective, K. Zhang, Z. Xu, A. Rosenkranz, Y. Song, T. Xue, Surface- and Tip-Enhanced Raman Scattering in, (2019) 1–16.

[43] Q. Li, C. Lee, R.W. Carpick, J. Hone, Substrate effect on thickness-dependent friction on graphene, Phys. Status Solidi Basic Res. 247 (2010) 2909–2914. https://doi.org/10.1002/pssb.201000555.

[44] D.S. Grierson, R.W. Carpick, Nanotribology of carbon-based materials, Nano Today. 2 (2007) 12–21. https://doi.org/10.1016/S1748-0132(07)70139-1.

[45] R.W. Carpick, M. Salmeron, Scratching the surface: Fundamental investigations of tribology with atomic force microscopy, Chem. Rev. 97 (1997) 1163–1194. https://doi.org/10.1021/cr960068q.

[46] U. Rabe, S. Amelio, M. Kopycinska, S. Hirsekorn, M. Kempf, M. Göken, W. Arnold, Imaging and measurement of local mechanical material properties by atomic force acoustic microscopy, Surf. Interface Anal. 33 (2002) 65–70. https://doi.org/10.1002/sia.1163.

[47] S.R. Polaki, K. Ganesan, S.K. Srivastava, M. Kamruddin, A.K. Tyagi, The role of substrate bias and nitrogen doping on the structural evolution and local elastic modulus of diamond-like carbon films, J. Phys. D. Appl. Phys. 50 (2017). https://doi.org/10.1088/1361-6463/aa6492.

[48] S.R. Polaki, N. Kumar, K. Ganesan, K. Madapu, A. Bahuguna, M. Kamruddin, S. Dash, A.K. Tyagi, Tribological behavior of hydrogenated DLC film: Chemical and physical transformations at nano-scale, Wear. 338–339 (2015) 105–113. https://doi.org/10.1016/j.wear.2015.05.013.

[49] L. Liu, M. Zhou, L. Jin, L. Li, Y. Mo, G. Su, X. Li, H. Zhu, Y. Tian, Recent advances in friction and lubrication of graphene and other 2D materials: Mechanisms and applications, Friction. 7 (2019) 199–216. https://doi.org/10.1007/s40544-019-0268-4.

[50] E. Riedo, F. Lévy, H. Brune, Kinetics of capillary condensation in nanoscopic sliding friction, Phys. Rev. Lett. 88 (2002) 1855051–1855054. https://doi.org/10.1103/physrevlett.88.185505.

[51] Z. yang Li, W. jing Yang, Y. ping Wu, S. bo Wu, Z. bing Cai, Role of humidity in

reducing the friction of graphene layers on textured surfaces, Appl. Surf. Sci. 403 (2017) 362–370. https://doi.org/10.1016/j.apsusc.2017.01.226.

[52] A. V. Sumant, D.S. Grierson, J.E. Gerbi, J. Birrell, U.D. Lanke, O. Auciello, J.A. Carlisle, R.W. Carpick, Toward the ultimate tribological interface: Surface chemistry and nanotribology of ultrananocrystalline diamond, Adv. Mater. 17 (2005) 1039–1045. https://doi.org/10.1002/adma.200401264.

[53] X. Wei, Z. Meng, L. Ruiz, W. Xia, C. Lee, J.W. Kysar, J.C. Hone, S. Keten, H.D. Espinosa, Recoverable Slippage Mechanism in Multilayer Graphene Leads to Repeatable Energy Dissipation, ACS Nano. 10 (2016) 1820–1828. https://doi.org/10.1021/acsnano.5b04939.

[54] C. Androulidakis, E.N. Koukaras, M. Hadjinicolaou, C. Galiotis, Non-Eulerian behavior of graphitic materials under compression, Carbon N. Y. 138 (2018) 227–233. https://doi.org/10.1016/j.carbon.2018.06.011.

[55] Y. You, Z. Ni, T. Yu, Z. Shen, Edge chirality determination of graphene by Raman spectroscopy, Appl. Phys. Lett. 93 (2008) 91–94. https://doi.org/10.1063/1.3005599.